# Integral Value Transformations: A Class of Discrete Dynamical Systems

*Sk. S. Hassan[1], A. Roy[1], P. Pal. Choudhury[1], B. K. Nayak[2]*

[1]*Applied Statistics Unit, Indian Statistical Institute, Calcutta 700108, India*

[2]*P. G. Department of Mathematics, Bhubaneswar 751004, India*

*Emails: sarimif@gmail.com, ananyaaroy1@gmail.com, pabitrapalchoudhury@gmail.com and bknatuu@yahoo.co.uk*

### Abstract

*Integral Value Transformations (IVTs) is a class of continuous maps in a discrete space $\mathbb{N}_0 = \mathbb{N} \cup \{0\}$. In this paper, these IVTs are considered to be Discrete Dynamical System maps in the space $\mathbb{N}_0$. The dynamical behaviour of the one dimensional non-linear autonomous iterative scheme of IVTs is deciphered through the light of Topological Dynamics.*

## 1. Introduction:

Integral Value Transformations was introduced by *Sk. S. Hassan* et al [1, 2, and 3] during 2009-10. In this section a brief description of the *Integral Value Transformations* (IVTs) is given as follows:

***Definition 1.1:*** A p-adic, k-dimensional, Integral Value Transformation is denoted by $IVT^{p,k}{}_j$. $IVT^{p,k}{}_j$ is a mapping from $\mathbb{N}_0{}^K$ to $\mathbb{N}_0$ as defined below.

$IVT^{p,k}{}_j : \mathbb{N}_0{}^K \rightarrow \mathbb{N}_0$

$IVT^{p,k}{}_j(n_1, n_2, \dots n_k) = (f_j(a_0{}^{n_1}, a_0{}^{n_2}, \dots, a_0{}^{n_k})\ f_j(a_1{}^{n_1}, a_1{}^{n_2}, \dots, a_1{}^{n_k}) \dots \dots f_j(a_{l-1}{}^{n_1}, a_{l-1}{}^{n_2}, \dots, a_{l-1}{}^{n_k}))_p = m$

where $n_1 = (a_0{}^{n_1} a_1{}^{n_1} \dots a_{l-1}{}^{n_1})_p$, $n_2 = (a_0{}^{n_2} a_1{}^{n_2} \dots a_{l-1}{}^{n_2})_p, \dots\dots n_k = (a_0{}^{n_k} a_1{}^{n_k} \dots a_{l-1}{}^{n_k})_p$

$f_j : \{0, 1, 2, \dots, p-1\}^k \rightarrow \{0, 1, 2, \dots, p-1\}$.

m is the decimal conversion from the p adic number.

Let us fix the domain of IVTs as $\mathbb{N}_0$ (k=1) and thus the above definition boils down to the following:

$$IVT^{p,1}{}_j(x) = \left(f_j(x_n)\ f_j(x_{n-1}) \dots \dots \dots f_j(x_1)\right)_p = m$$

where m is the decimal conversion from the p adic number, and $x = (x_n\ x_{n-1} \dots \dots x_1)_p$.

Let $T^{p,1}$ denotes the set of all one-dimensional p-adic Integral Value Transformations.

The definition is illustrated below:

For p=3, k=1

| Variables | $f_7$ | $f_{16}$ |
|---|---|---|
| 0 | 1 | 1 |
| 1 | 2 | 2 |
| 2 | 0 | 1 |

$x = 55 = (2001)_3$

$IVT^{3,1}{}_7(x) = (f_7(2)\, f_7(0)\, f_7(0)\, f_7(1))_3 = (0112)_3 = 14$

$IVT^{3,1}{}_{16}(x) = (f_{16}(2)\, f_{16}(0)\, f_{16}(0)\, f_{16}(1))_3 = (1112)_3 = 41$

***Definition 1.2:*** A function f is said to be a *Collatz-like function* if $\exists\, m \in \mathbb{N}_0$ such that $f^m(n) = c$, c is a fixed point over the iterations for any choice of n.

In any p-adic system, it has been seen that there are $(p^{p-1} - 1)$ number of Collatz-like functions [4].

IVTs have been studied on similar lines as Cellular automata, introduced by Von Neumann. Cellular automata have been studied extensively over the past forty years since its inception by computer scientists and mathematicians alike and yielded many beautiful results. The characterisation of Cellular automata by Hedlund [5] provided a new dimension of looking at these functions and thus started the endeavours of studying Cellular automata through topological dynamics. .

***Definition 1.3***: A semi-group $(G, f)$ acting on a space M is called a *dynamical system* if a mapping

T: G x M $\to$ M defined as T (g, x) = $T_g(x)$ such that $T_{f(g,h)} = f(T_g, T_h)$ . Further, if $G = \mathbb{N}_0$ or

G = Z, then the system is called a *Discrete Dynamical System [DDS] [5, 6]*.

IVTs form a *discrete dynamical system* when applied iteratively and this opens up a vast unexplored area. It is interesting to see how these functions evolve over time, form chaotic patterns, etc. The real motivation is to make an attempt at understanding how these IVTs evolve over time. Dynamical systems could throw some light in this aspect thereby aiding us in comprehending the time evolution of these functions. To meet this end, we first define a dynamical system of IVTs which is done in section 3 along with a few results.

## 2. IVTs as Discrete Dynamical Systems

***Theorem-2.1***: $(\mathbb{N}_0, T)$ is a discrete dynamical system. The function T is defined as

T: $\mathbb{N}_0 \times \mathbb{N}_0 \to \mathbb{N}_0$ as $T(n, x) = T_n(x)$ Where

$T_n(x) = (IVT^{p,1}{}_j)^n(x) = \underbrace{[(IVT^{p,1}{}_j)\,(IVT^{p,1}{}_j)\ldots\ldots(IVT^{p,1}{}_j)]}_{n\ \text{times}}\,(x)$

*Proof:* Clearly, $(\mathbb{N}_0, x)$ is a semi-group acting on the space $\mathbb{N}_0$.

Now,

$T_{n_1 \times n_2} = (IVT^{p,1}{}_j)^{n_1 \times n_2} = \underbrace{[(IVT^{p,1}{}_j)\,(IVT^{p,1}{}_j)\ldots\ldots(IVT^{p,1}{}_j)]}_{n_1 \times n_2\ times}$

$$= \underbrace{[(IVT^{p,1}_j)(IVT^{p,1}_j)\ldots\ldots(IVT^{p,1}_j)]}_{n_1\ times} \times \underbrace{[(IVT^{p,1}_j)(IVT^{p,1}_j)\ldots\ldots(IVT^{p,1}_j)]}_{n_2\ times}$$

$$= T_{n_1} \times T_{n_2}$$

[Remark: by another definition,

$T(0, x) = (IVT^{p,1}_j)^0(x) = x$

$T(n_2, T(n_1, x)) = T(n_2, (IVT^{p,1}_j)^{n_1}(x)) = (IVT^{p,1}_j)^{n_2}[(IVT^{p,1}_j)^{n_1}(x))] = (IVT^{p,1}_j)^{n_1 \times n_2}(x)) =$

$T(n_2 \times n_1, x)$ for all $n_1, n_2 \in I(x) = \{n \in \mathbb{N}_0 \mid (n,x) \in \mathbb{N}_0 \times \mathbb{N}_0\} = \mathbb{N}_0$, $x \in \mathbb{N}_0$.]

Therefore, $(\mathbb{N}_0, T)$ is a discrete dynamical system.

Here $T(n, x)$ is the evolution function and n is the evolution parameter of the dynamical system and $X_0$ is the initial state/condition and $\mathbb{N}_0$ is the state/phase space.

Corresponding to each p and j, we get a different dynamical system.

For a fix variable n, then $T_n = (IVT^{p,1}_j)^n : \mathbb{N}_0 \to \mathbb{N}_0$ is called the *flow through x*.

Further, if $IVT^{p,1}_j$ is an invertible map, then the dynamical system is *invertible*.

***Definition 2.1***: The orbit of $x_0$ is a set of points denoted by

$\gamma(x_0) = \{x_n \mid n \in \mathbb{N}_0, x = x_0, x_{n+1} = (IVT^{p,1}_j)(x_n)\}$

$= \{x_0, (IVT^{p,1}_j)(x_0), (IVT^{p,1}_j)^2(x_0), (IVT^{p,1}_j)^3(x_0), \ldots\ldots\ldots\}$

***Definition 2.2***: A point p is called a *fixed point* of a function f if $f(p) = p$ and the set of fixed points is denoted by Fix (f).

For any $IVT^{p,1}_j$, $x = (x_n\ x_{n-1}\ldots\ldots x_1)_p$ will be a fixed point of $IVT^{p,1}_j$

i.e. $IVT^{p,1}_j(x) = \left(f_j(x_n)\ f_j(x_{n-1})\ldots\ldots f_j(x_1)\right)_p = (x_n\ x_{n-1}\ldots\ldots x_1)_p = x$ if $f_j(x_i) = x_i$ for each $i=1,2,\ldots,n$.

***Definition 2.3***: A fixed point of $f^n$ is called a periodic point of period n of f and the set of periodic pints is denoted by Per (f).

Example: We know $IVT^{2,1}_1$ is a Collatz like function so we expect its orbit around any point to be finite and containing 0. Its orbit around $x_0$ is given by

$\gamma(x_0) = \{x_0, (IVT^{2,1}_1)(x_0), (IVT^{2,1}_1)^2(x_0), (IVT^{2,1}_1)^3(x_0), \ldots\ldots\ldots\}$
$\gamma(0) = \{0,1\}$, $\gamma(1) = \{0,1\}$,
$\gamma(2) = \{2, (IVT^{2,1}_1)(2), (IVT^{2,1}_1)^2(2), (IVT^{2,1}_1)^3(x_0), \ldots\ldots\} = \{2,1,0\}$, $\gamma(3) = \{3,1,0\}$,
$\gamma(4) = \{4,3,1,0\}$, $\gamma(5) = \{5,2,1,0\}$, $\gamma(6) = \{6,1,0\}$, $\gamma(7) = \{7,1,0\}$, $\gamma(15) = \{15,1,0\}$

It's worth noting that the orbits of the Merseene numbers are 3-point sets which consists of the number itself and 0 and 1 in case of $IVT^{2,1}_1$. The orbits basically give the path/trajectory of convergence of the Collatz like functions.

For $IVT^{2,1}_2(x) = x$, $\gamma(x_0) = \{x_0, (IVT^{2,1}_2)(x_0), (IVT^{2,1}_2)^2(x_0), (IVT^{2,1}_2)^3(x_0), \ldots\ldots\ldots\} = \{x_0\}$

The orbit of any steady state equilibrium/fixed point will be the point itself. For $IVT^{2,1}{}_3(x)$,

$\gamma(x_0) = \{ x_0, (IVT^{2,1}{}_3)(x_0), (IVT^{2,1}{}_3)^2(x_0), (IVT^{2,1}{}_3)^3(x_0), \ldots \ldots \}$

It is possible to classify the orbits in the following categories of steady state equilibria/fixed points, periodic points and non-periodic points.

If $IVT^{p,1}{}_j$ is periodic of period n, then the orbit of $x_0$ is

$\gamma(x_0) = \{ x_0, (IVT^{p,1}{}_j)(x_0), (IVT^{p,1}{}_j)^2(x_0), (IVT^{p,1}{}_j)^3(x_0), \ldots \ldots \ldots, (IVT^{p,1}{}_j)^{p-1}(x_0) \}$

Finding the set of fixed points itself is an arduous task in the first place owing to the complexity of the functions. We will first look at the set of fixed points in a particular case and then make an attempt towards a generalisation.

## 3. *Non-Linear System of IVT*

The dynamical system of IVTs forms a Non-linear system and here it has been exploited the existing literature in this area to analyse the stability of fixed points.

$T : \mathbb{N}_0 \times \mathbb{N}_0 \to \mathbb{N}_0$ as

$T(n, x) = T_n(x)$

where $T_n(x) = (IVT^{p,1}{}_j)^n(x) = \underbrace{[(IVT^{p,1}{}_j)(IVT^{p,1}{}_j) \ldots \ldots (IVT^{p,1}{}_j)]}_{n\ times}(x)$

Let $x_0$ be the initial condition.

$T_0(x_0) = (IVT^{p,1}{}_j)^0(x_0) = x_0$

$x_1 = T_1(x_0) = (IVT^{p,1}{}_j)^1(x_0)$, $x_2 = T_2(x_0) = (IVT^{p,1}{}_j)^2(x_0)$, $x_3 = T_3(x_0) = (IVT^{p,1}{}_j)^3(x_0)$, ……

$\{x_n\}$ gives us the trajectory of the non-linear system and the iterative scheme is given by

$x_n = IVT^{p,1}{}_j(x_{n-1})$.

***Definition 3.1***: A steady state equilibrium of the equation $x_n = IVT^{p,1}{}_j(x_{n-1})$ is a point $\bar{x} \in \mathbb{N}_0$ such that $IVT^{p,1}{}_j(\bar{x}) = \bar{x}$ that is $\bar{x}$ is a fixed point of $IVT^{p,1}{}_j$.

Stability Analysis of steady state equilibria of discrete dynamical systems is based on some propositions and/or explicit solution of the non-linear, autonomous (the parameters/coefficients *a* and *b* in the difference equation $x_n = ax_{n-1} + b$ are independent of time), one-dimensional dynamical systems after reducing the non-linear system to a linear system.

A linear system is called *locally stable* if for a small perturbation to the system, it converges asymptotically to the original equilibrium. A linear system is called *globally stable* if irrespective of the extent of perturbation, it converges asymptotically to the original equilibrium. Mathematically, the definition is as follows:

***Definition 3.2***: A steady state equilibrium $\bar{x}$, of the linear difference equation $x_n = ax_{n-1} + b$ is called *globally (asymptotically) stable* if $\lim_{n \to \infty} x_n = \bar{x}$ for all $x_0 \in \mathbb{N}_0$ and is called *locally (asymptotically) stable* if there exists $r > 0$ such that $\lim_{n \to \infty} x_n = \bar{x}$ for all $x_0 \in N(\bar{x}, r)$, a neighbourhood of the point $\bar{x}$.

Now, for the non-linear system $x_n = IVT^{p,1}{}_j(x_{n-1})$ …… (1)

Through the Taylor series expansion of the non-linear system around the fixed point $\bar{x}$ of (1), it can be reduced to a linear system around the steady state equilibrium $\bar{x}$ to approximate the behaviour around the fixed point by a linear system. The Taylor series expansion is the following

$$x_{n+1} = IVT^{p,1}{}_j(x_n) = IVT^{p,1}{}_j(\bar{x}) + (x_n - \bar{x})DIVT^{p,1}{}_j(\bar{x}) + \frac{(x_n - \bar{x})^2}{2!}$$

$$D^2 IVT^{p,1}{}_j(\bar{x}) + \ldots\ldots + \frac{(x_n - \bar{x})^k}{k!} D^k IVT^{p,1}{}_j(\bar{x}) + \ldots\ldots$$

where $D^k IVT^{p,1}{}_j(\bar{x})$ denotes the kth derivative of $IVT^{p,1}{}_j$ at the point $\bar{x}$. The concept of derivative in $T^{p,1}$ is defined in [3]

The linearized system around the fixed point is

$$x_{n+1} = IVT^{p,1}{}_j(\bar{x}) + (x_n - \bar{x})DIVT^{p,1}{}_j(\bar{x}) \text{ neglecting the higher order terms}$$

$$= [DIVT^{p,1}{}_j(\bar{x})]x_n + [IVT^{p,1}{}_j(\bar{x}) - \bar{x}DIVT^{p,1}{}_j(\bar{x})]$$

$$= ax_{n-1} + b \quad\ldots\ldots(2)$$

where $a = DIVT^{p,1}{}_j(\bar{x})$ and $b = [IVT^{p,1}{}_j(\bar{x}) - \bar{x}DIVT^{p,1}{}_j(\bar{x})]$

By a proposition stated below

The linearized system (2) is globally stable iff $|a| = |DIVT^{p,1}{}_j(\bar{x})| < 1$

That is, for any $x_0 \neq \bar{x}, \lim_{n\to\infty} x_n = \bar{x}$ if $|a| = |DIVT^{p,1}{}_j(\bar{x})| < 1$ and convergence is monotonic if 0<a<1 and oscillatory if -1<a<0

[*Proposition*: A necessary and sufficient condition for global stability of a linear system

***Definition 3.3:*** A steady state equilibrium of the difference equation $x_n = ax_{n-1} + b$ is globally stable if and only if $|a| < 1$. Further for $x_0 \neq \bar{x}, \lim_{n\to\infty} x_n = \bar{x}$ if $|a| < 1$, convergence is monotonic if $0 < a < 1$ and oscillatory if $-1 < a < 0$]

Since the above linear system (2) is obtained by linearizing the non-linear system around $\bar{x}$ (i.e; in a neighbourhood of $\bar{x}$), therefore the global stability of the linear system (2) ensures only the local stability of the non-linear system (1).

Thus, the dynamical system (1) is locally stable around the steady state equilibrium/fixed point $\bar{x}$ iff $|DIVT^{p,1}{}_j(\bar{x})| < 1$.

Thus we have established the condition for local stability of the non-linear system (1) around the steady state equilibrium $\bar{x}$.

For deriving a condition for the global stability of unique steady state equilibrium, we will invoke the contraction mapping theorem/ Banach's fixed point theorem. To do this, we first define a metric d on $\mathbb{N}_0$ such that $(\mathbb{N}_0, d)$ is a complete metric space and find all the contraction mappings for some p and j in a p-adic system. Now we apply the fixed point theorem to arrive at a unique fixed point (which ensures the global stability of the non-linear system).

*Banach's Fixed Point Theorem*:

*If (X, d) is a complete metric space and f: X→X is a contraction mapping, then f will have a unique fixed point where f: X→X is a contraction mapping means d (f(x), f(y)) ≤ λ d(x, y) for all x, y Є X and for some λ Є (0, 1).*

## 4. Topological dynamics of IVTs

Since the domain of the Integral Value Transformations is $\mathbb{N}_0$, it has to be first endowed with a topological and measure theoretic structure. Thus, let $\mathbb{N}_0$ be given the discrete topology (the co-countable topology also works*) denoted by $\tau_d$ which is essentially the collection of all subsets of $\mathbb{N}_0$ (with cardinality $\aleph_0$ i.e; aleph – nul). Thus $\tau_d$ being the power set of $\mathbb{N}_0$ has cardinality $2^{\aleph_0}$=c, referred to as the cardinality of the continuum.

This topology induces the discrete metric given by

$d : \mathbb{N}_0 \times \mathbb{N}_0 \to \mathbb{R}^+$

$d(x, y) = \begin{cases} 1, & x \neq y \\ 0, & x = y \end{cases}$

Thus, all the subsets of $\mathbb{N}_0$ are clopen. Now we give $\mathbb{N}_0$ a measure theoretic treatment.

Let $(\mathbb{N}_0, ß, \mu)$ denote the measure space on $\mathbb{N}_0$ where ß is the σ- algebra on $\mathbb{N}_0$ and μ is the counting measure defined on $\mathbb{N}_0$ given below:

$\mu : ß \to \mathbb{R}^+$

$\mu(A) = \begin{cases} |A|, & \text{if } A \text{ is finite} \\ \infty, & \text{if } A \text{ is infinite} \end{cases}$

where |A| denotes the cardinality of A.

Now, it is a position to delve further into the topological properties of the IVTs.

The very fundamental and basic question can be raised are the IVTs measure preserving transformations?

The answer is not always affirmative.

Let $A \in ß$,

$(IVT^{p,1}_j)^{-1}(A) = \{x \in \mathbb{N}_0 \text{ such that } (IVT^{p,1}_j)(x) \in A\}$

If A is finite and $IVT^{p,1}_j$ is bijective (and not Collatz like), then $\mu(A) = \mu\left((IVT^{p,1}_j)^{-1}(A)\right)$

If A is infinite and $IVT^{p,1}_j$ is neither the zero function nor the Collatz like functions, then too $\mu(A) = \mu\left((IVT^{p,1}_j)^{-1}(A)\right)$.

Thus, in the above cases, $IVT^{p,1}_j$ is measure preserving and hence $(IVT^{p,1}_j)^n$ will also be measure preserving. Thus, for the dynamical system given below:

$T: \mathbb{N}_0 \times \mathbb{N}_0 \to \mathbb{N}_0$ as

$T(n, x) = T_n(x)$

Where $T_n(x) = (IVT^{p,1}_j)^n(x) = \underbrace{[(IVT^{p,1}_j)(IVT^{p,1}_j)\ldots\ldots(IVT^{p,1}_j)]}_{n \text{ times}}(x)$

T is also *measure preserving*.

T is *ergodic* since there does not exist a set of measure strictly between 0 and 1 such that $T^{-1}(A) = A$.

Now let us look at a few examples.

Consider the Collatz-like function $IVT^{2,1}_1$ form $T^{2,1}$

Let $A = \{0\}$.

Then $(IVT^{p,1}_j)^{-1}(A) = \{x \in \mathbb{N}_0 \text{ such that } (IVT^{p,1}_j)(x) \in A\} = \{0,3,7,15,31,\ldots\ldots\} = \{2^n - 1 : n = 1,2,3,\ldots\}$ that is the set of all merseene numbers.

Clearly, $IVT^{2,1}_1$ is not measure preserving as $\mu(A) = 1$ but $\mu\left((IVT^{p,1}_j)^{-1}(A)\right) \to \infty$

Let A be any finite or infinite set. In the 2-adic system, $IVT^{2,1}_2$ is the identity function, under which the inverse image of any finite or infinite set is the set itself and hence it is measure preserving.

In the 3-adic system, consider the function $IVT^{3,1}_{15}$ which is bijective but not Collatz like.

Let $A = \{0, 1, 2, 7, 10\}$. $(IVT^{3,1}_{15})^{-1}(A) = \{0, 2, 1, 5, 20\}$ which has the same measure as A and thus $IVT^{3,1}_{15}$ is a measure preserving transformation.

### *4.1.1 Topological Conjugacy of two IVT-dynamical systems*

***Definition 4.1:*** If (X, f) and (Y, g) are two dynamical systems, then (Y, g) is called a factor of (X, f) if there exists a continuous surjection h: $X \to Y$ such that $h \cdot f = g \cdot h$. If the function h is also a homeomorphism, then the two systems are called ***topologically conjugate***.

For any $IVT^{p,1}_{j_1}$ and $IVT^{p,1}_{j_2}$, the dynamical systems ($\mathbb{N}_0$, $IVT^{p,1}_{j_1} \cdot IVT^{p,1}_{j_2}$) and ($\mathbb{N}_0$, $IVT^{p,1}_{j_2} \cdot IVT^{p,1}_{j_1}$) will be factors of each other since

$IVT^{p,1}_{j_1} \cdot (IVT^{p,1}_{j_2} \cdot IVT^{p,1}_{j_1}) = (IVT^{p,1}_{j_1} \cdot IVT^{p,1}_{j_2}) \cdot IVT^{p,1}_{j_1}$ and

$IVT^{p,1}_{j_2} \cdot (IVT^{p,1}_{j_1} \cdot IVT^{p,1}_{j_2}) = (IVT^{p,1}_2 \cdot IVT^{p,1}_{j_1}) \cdot IVT^{p,1}_{j_2}$ where $IVT^{p,1}_{j_1}$ and $IVT^{p,1}_{j_2}$ are continuous surjections.

($\mathbb{N}_0$, $IVT^{3,1}_{13}$) is a factor of ($\mathbb{N}_0$, $IVT^{3,1}_{18}$) since

$IVT^{3,1}_{16} \cdot IVT^{3,1}_{18} = IVT^{3,1}_{13} \cdot IVT^{3,1}_{16} (= IVT^{3,1}_{13})$ where $IVT^{3,1}_{16}$ is a continuous surjection.

Now let us define a mapping h: $\mathbb{N}_0 \to \mathbb{N}_0$ as

$h(x) = h((x_n\, x_{n-1} \ldots\ldots x_1)_p) = ((x_n \oplus_p 1)\, (x_{n-1} \oplus_p 1) \ldots\ldots (x_1 \oplus_p 1))_p$.

Clearly, h is a homeomorphism. ($\mathbb{N}_0$, $IVT^{3,1}_{16}$) and ($\mathbb{N}_0$, $IVT^{3,1}_8$) are topologically conjugate dynamical systems as $h \cdot IVT^{3,1}_{16} = IVT^{3,1}_8 \cdot h$. Thus these two systems show the same dynamical behaviour.

This is a trivial way of looking at the dynamics of these functions and may pose some limitations in the future regarding the analysis of these functions. To overcome this shortcoming, we adopt another way of looking at the topological dynamics given below.

We translate the dynamical system give in section 2 to an analogous dynamical system with the hope that this will provide us with a simpler system. To do this, we express the Integral value transformation as a composition of three maps and look at the dynamics of this function through symbolic dynamics.

$IVT^{p,1}_j : \mathbb{N}_0 \to \mathbb{N}_0$

$IVT^{p,1}_j = C^{-1}FC$ where

$C: \mathbb{N}_0 \to \bigcup_{n \in \mathbb{N}_0} B_p^n = B_p^\mathbb{N}$ denotes the function for conversion of a natural number into its p-adic representation given by $C(m) = (x_0, x_1, x_2, \ldots, x_{k-1})$ where $x_0 = m\%p$, $y_0 = \frac{m}{p} + x_0$, $x_1 = y_0\%p$, and so on.

$F: B_p^k \to B_p^k$ is defined as $F((x_0, x_1, x_2, \ldots, x_{k-1})) = (f_j(x_0), f_j(x_1), f_j(x_2), \ldots, f_j(x_{k-1}))$ for some $f_j: B_p \to B_p$.

$C^{-1}: \bigcup_{n \in \mathbb{N}_0} B_p^n \to \mathbb{N}_0$ denotes the conversion from a p-adic number to a natural number given by $C^{-1}((x_0, x_1, x_2, \ldots, x_{k-1})) = \sum_{i=0}^{k-1} x_i p^{k-1-i} = m$ (natural number)

Thus, $(IVT^{p,1}_j)^n(x) = \underbrace{[(IVT^{p,1}_j)(IVT^{p,1}_j)\ldots\ldots(IVT^{p,1}_j)]}_{n \text{ times}}(x) = (C^{-1}FC)^n = (C^{-1}F^nC)$

Where $F^n: B_p^k \to B_p^k$ such that $F^n((x_0, x_1, x_2, \ldots, x_{k-1})) = ((f_j)^n(x_0), (f_j)^n(x_1), (f_j)^n(x_2), \ldots, (f_j)^n(x_{k-1}))$

## *5. Conclusion*

Thus, so far we have examined some characteristics of Collatz-like functions and a few non-Collatz-like functions and stability analysis of fixed points from a theoretical point of view but only for derivable IVTs. We'll try to extend this study the same for all IVTs. We have also introduced some basics of topological dynamical systems and are in the process of delving deeper into this.

*Acknowledgement:* The authors express their gratitude to Dr. B. S. Daya Sagar and Dr. Sudhakar Sahoo for their thoughtful research comments and suggestions.